\documentclass[aps,english,prl,twocolumn,superscriptaddress]{revtex4}
\usepackage{graphicx}

\begin{document}

\title{Dissociative Adsorption of Molecules on Graphene and Silicene}

\author{H. Hakan G\" urel}
\affiliation{UNAM-National Nanotechnology Research Center, Bilkent University, 06800 Ankara, Turkey}
\affiliation{Institute of Materials Science and Nanotechnology, Bilkent University, Ankara 06800, Turkey}
\affiliation{Department of Information Systems Engineering, Kocaeli University, Kocaeli 41380, Turkey}
\author{V. Ongun \"{O}z\c{c}elik}
\affiliation{UNAM-National Nanotechnology Research Center, Bilkent University, 06800 Ankara, Turkey}
\affiliation{Institute of Materials Science and Nanotechnology, Bilkent University, Ankara 06800, Turkey}
\author{S. Ciraci}
\affiliation{Department of Physics, Bilkent University, Ankara 06800, Turkey}

\begin{abstract}
We study the interaction of H$_2$, O$_2$, CO, H$_2$O and OH molecules with the vacancy defects of graphene and silicene. Atoms around the bare vacancy reconstruct and specific chemically active sites are created. While H$_2$, O$_2$ and CO remain intact on both pristine graphene and silicene, these molecules can dissociate when they are placed at the close proximity of these chemically active sites and nucleate centers for the hydrogenation and oxygenation. Saturation of the dangling bonds at the defect sites by constituent atoms of dissociated molecules gives rise to significant modification of electronic and magnetic properties. We analyzed the mechanism of the dissociation and revealed a concerted action of surrounding host atoms together with dissociated molecules to lower the energy barrier needed for dissociation. The dissociations of H$_2$O and OH are hindered by high energy barriers. Our study suggests that graphene and silicene can be functionalized by creating meshes of single vacancy, where specific molecules can dissociate, while some other molecules can be pinned.

\end{abstract}

\maketitle

\section{Introduction}
Recent studies in nanoscale physics have aimed at discovering new monolayer materials and revealing their properties under different conditions. Among these materials, graphene which is a single layer honeycomb structure of carbon, has attracted great interest in various fields ranging from electronics to biotechnology due to its unique properties such as high mechanical strength, chemical stability and exceptional ballistic conductance.\cite{novoselov2004,geim2007,zhang} Similarly, silicene, a graphene like allotrope of silicon was also shown to be stable when alternating atoms of hexagons are buckled.\cite{seymur2009,engin} More recently single and multilayer silicene and its derivatives have been grown on Ag(111) substrate \cite{nakamura,lay2010,lay2012,seymur3,seymur4}.

Although both theoretical and experimental studies \cite{yazyev2010,hashimoto2004,krash2007,crespi1996,kim2011,singh2009,faccio2010} support the stability of graphene and silicene, local defects can always exist at finite temperatures. Among these, the most frequently observed defect type is the single vacancy in single layer honeycomb structures. Vacancy defects in graphene usually emerge during epitaxial growth at grain boundaries or step edges and they alter the properties of the pristine structures significantly.\cite{neto2009,reina2008,banhart2011,growth} They create localized or resonance defect states, modify the magnetic ground state and distort the pristine, single layer honeycomb structure. Physical methods such as stress, irradiation, and sublimation can also result in a non-equilibrium concentration of such defects. Vacancy defects which are generated for various reasons, may either stay permanent or they may be self-healed under proper external conditions. It was recently observed that graphene self-heals its defects if it is placed in a reservoir with enough host external atoms.\cite{zan2012, robertson2013,vozcelik2013}. The self-healing may result in perfect honeycomb structure or Stone-Wales\cite{stone86, wales98} type defects may be generated as a result of the healing process. Stone-Wales defects by themselves are healed by the rotation of specific a C-C bond.\cite{growth,wang} These observations have also been supported by theoretical studies and a similar self-healing mechanism was proposed for defected silicene \cite{vozcelik2013}.  Transmission electron microscopy observations and theoretical calculations of self healing are crucial for future graphene and silicene based device applications, such as sensors, because the properties of the pristine structures, especially the conductance undergo dramatic changes after defect formation and also after the healing process.

Even though formation of defects appears to be undesirable at the first sight, vacancy defects, large holes and their meshes can also be created willingly with the help of external agents in order to modify the properties of graphene and silicene. Carbon and Si atoms surrounding the vacancy or hole have lower coordination and hence have dangling bonds which increase the local chemical activity.\cite{boukhvalov,carlsson} Incidently, not only vacancies, but also inhomogeneities in the charge distribution induced by wrinkles on free-standing graphene\cite{glukhova} or by radial deformations on carbon nanotubes\cite{swnt} were shown to enhance chemical activity locally.

Owing to the increased chemical activity,  single foreign atoms can adsorb at the defected site. Even specific molecules can dissociate into constituent atoms, which, in turn, saturate the host atoms with lower coordination \cite{lin2007,lin2011,can,nardelli,jiang2013}. In particular, the dissociation of H$_{2}$O and H$_2$ is of crucial importance for hydrogen economy,\cite{sefa,taner,engin2,can2} even though the interaction of these molecules with pristine graphene and silicene are rather weak. The oxidation of graphene followed by the adsorption of O adatoms is of particular interest, because an almost reversible oxidation-deoxidation mechanism leading to a continuous metal-insulator transition is possible.\cite{yao,dikin,wei,topsakal} The interaction between O$_2$ and pristine graphene or silicene is very weak, but the adsorption of O atoms at the vacancy sites following the dissociation of O$_2$ appears to be a precursor of oxidation of these single layer honeycomb structures.\cite{silicatene} The interaction of CO molecule with graphene has been also a subject of interest for fabrication of gas sensors or filters.\cite{tanveer} Recently, the coverage of monolayer structures by physisorbed O$_2$ and H$_2$O molecules has been shown to lead also significant enhancements in optical and electronic properties.\cite{sef1,sef2}

In this paper we investigate the interaction of H$_2$, O$_2$, CO, H$_2$O, OH molecules with graphene and silicene at the single vacancy sites. We start with the analysis of the vacancy in the absence of external molecules and show how it reconstructs and forms chemically active sites. Then, we introduce these molecules above the reconstructed vacancy site at a distant height and perform structure optimization calculations for different heights to reveal their dissociation or bonding mechanisms. In this respect, this study is unique since the interaction between single vacancy site and various molecules are studied in a wide range of distances. We find that H$_2$, O$_2$ and CO molecules can dissociate around defects. For H$_2$ and O$_2$, their constituent O and H atoms are bonded readily to the atoms surrounding the vacancy. Upon the dissociation of CO molecule, the vacancy defects are healed and electronic as well as magnetic properties are dramatically modified. Under certain circumstances, CO can adsorb as a molecule without dissociation. On the other hand, H$_2$O and OH remain intact and stay attached to the vacancy site as a molecule.

\section{Method}
We performed first-principles calculations using spin polarized Density Functional Theory within generalized gradient approximation including van der Waals (vdW) corrections \cite{grimme06}. We used projector-augmented wave potentials \cite{paw}, and the exchange-correlation potential was approximated by Perdew, Burke and Ernzerhof (PBE) functional \cite{pbe}. Single vacancy is treated within the periodically repeating supercell method. Brilliouin zones of the (8$\times$8) supercells of graphene and silicene were sampled by 5x5x1 special \textbf{k}-points within the Monkhorst-Pack scheme.\cite{MP} where the convergence of total energies with rescpect to numbe rof k-points was tested. This corresponds to a rather fine grained sampling. A plane-wave basis set with energy cutoff value of 550 eV was used. For all structures and adsorption geometries, atomic positions and lattice constants were optimized by using the conjugate gradient method, where the total energy and atomic forces are minimized. The energy convergence value between two consecutive steps was chosen as $10^{-5}$eV.

\begin{figure}
\begin{center}
\includegraphics[width=8cm]{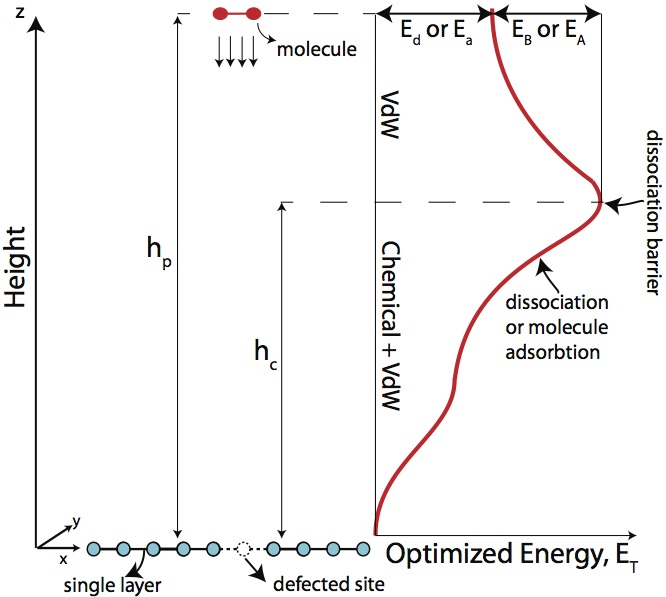}
\caption{A schematic description of the method used to investigate the interaction between a molecule described by red balls and defected single layer graphene or silicene shown by blue balls. Left: The approach of the molecule from distant height on the defect site vertically. The molecule may be trapped in a shallow minimum at $h_p$ and is bound by van der Waals interaction. The barrier to dissociation or molecular adsorption occurs at $h_c$. Right: The total energy of optimized structure of molecule+defected graphene or silicene versus the height of the molecule $z$. ($x,y$) plane is parallel to the single layer. Energy barrier for dissociation $E_B$ or molecular adsorption $E_A$, dissociation energy $E_d$ and molecular adsorption energy $E_a$ are indicated.}
\label{fig0}
\end{center}
\end{figure}

To investigate the interaction between H$_2$, O$_2$, CO, H$_2$O, OH molecules and defected graphene or silicene, we carried out the following calculations: First, the molecule is placed at a distant height ($z$) from the vacancy site for various fixed lateral ($x,y$) positions and subsequently they are lowered step by step by performing structure optimization through the minimization of the total energy and atomic forces. Initially, a molecule can be trapped or physisorbed in a shallow minimum at the height $h_p$ above the defect site, where it may become weakly bound by the vdW interactions or it may stay inert. As the molecule is brought closer to the vacancy site, there may be a barrier at a critical height $h_c$. If the molecule is further away from $h_c$, it goes away from the surface or becomes trapped at $h_p$. However, as soon as the molecule gets closer than $h_c$, it is attracted towards the defect site by  chemical interaction forces and moves even closer to dissociate or to adsorb as a molecule. We define the energy barrier for dissociation or molecular adsorption as the difference between the total energy
values of molecule (M)+defected graphene or silicene layer (L) optimized at $h_c$ and at $h_p$; namely,
$E_{B}=E_{T}[M+L;h=h_{c}]-E_{T}[M+L;h=h_p]$. $E_{B} >$ 0 indicates that there is barrier preventing the molecule
to approach the defected site in order to set up relatively stronger chemical bonds eventually leading to
dissociation or molecular adsorption. The dissociation energy is calculated as the difference between the total
energy of the molecule+defected layer optimized at $h_p$ and the total energy of the defected layer with constituent atoms (A) of molecules adsorbed at the defect site, i.e. $E_{d}=E_{T}[M+L;h_p]-E_{T}[A+L]$. In the case of molecular adsorption, the adsorption energy $E_a=E_{T}[M+L;h_p]-E_{T}[M+L]$. $E_{d} > $0 or $E_{a} > $0 implies that the dissociation or molecular adsorption is an exothermic process (See ~\ref{fig0}). It should be noted that $E_d$ differs from the dissociation energy of the free molecule into its constituent atoms. In all of the calculations periodic boundary conditions are used within the supercell geometry and the vacuum spacing between graphene or silicene layers in adjacent supercells is taken as 15 \AA. Numerical calculations were carried out using VASP software \cite{vasp}.

\begin{figure}
\begin{center}
\includegraphics[width=8cm]{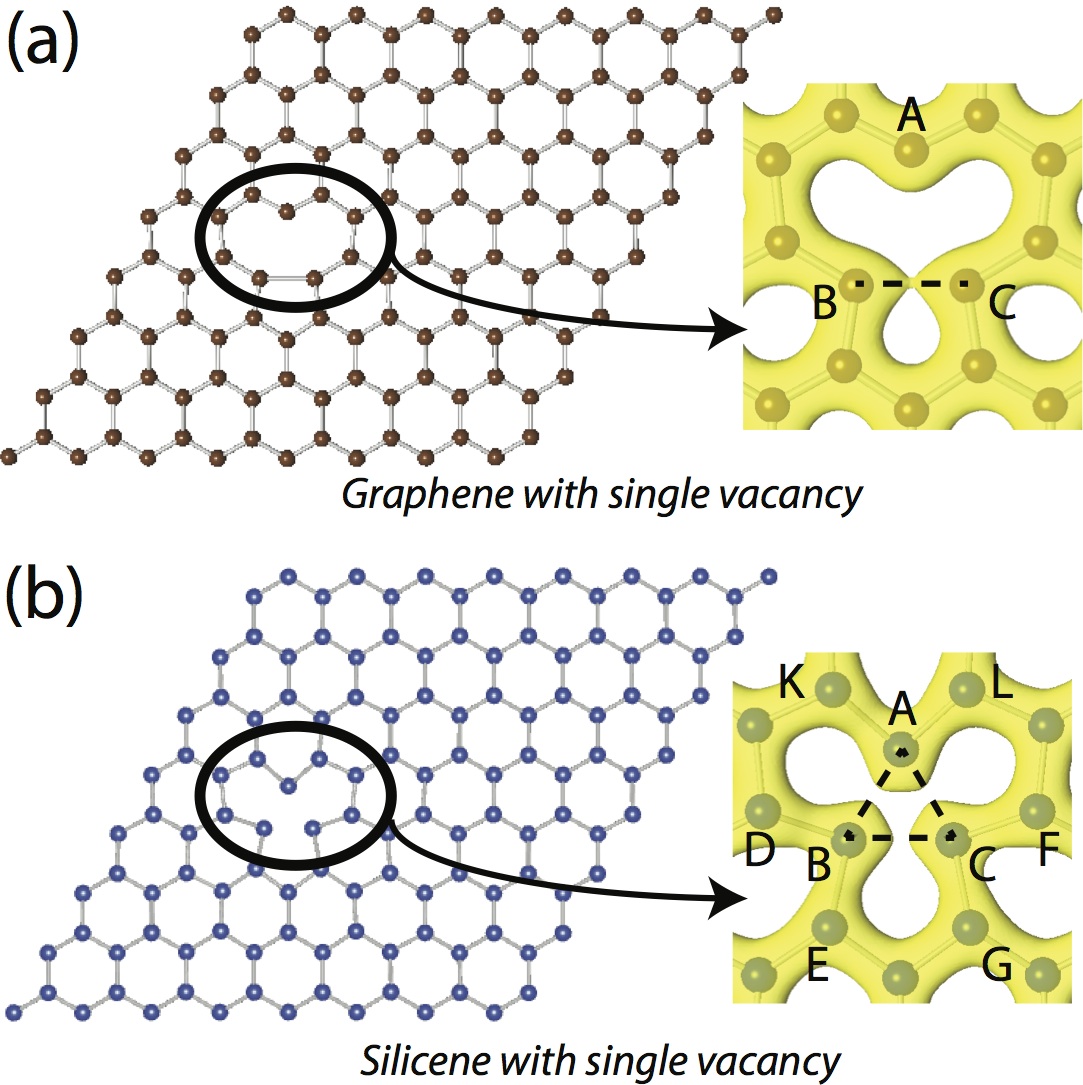}
\caption{Top view of the reconstructed, single vacancy defect in (a) graphene and (b) silicene. The vacancy reconstructs asymmetrically in graphene, and symmetrically in silicene. The distribution of total charge around the defect shows the chemically active sites. The dashed lines indicate the second nearest-neighbor distances (between atoms surrounding the vacancy, A,B and C) which are reduced upon reconstruction.}
\label{fig1}
\end{center}
\end{figure}

\section{Results and discussions}

\subsection{Reconstructions of single vacancies}
The reconstructions of single vacancies in graphene and silicene, which were observed in recent experimental works\cite{neto2009,reina2008,banhart2011} and treated in theoretical studies \cite{vozcelik2013}, establish the starting point of our study. To minimize vacancy-vacancy coupling, one single vacancy is created in each $(8 \times 8)$ supercell of pristine graphene and silicene. Subsequently, lattice constants and the atomic positions in the supercell were optimized using self-consistent field conjugate gradient method. The final reconstructed geometries are presented in ~\ref{fig1} for graphene and silicene, respectively. As the defected structures reconstruct, they shrink and the atoms reorient themselves to close the hole of the vacancy. However, these reconstructions are not sufficient to completely heal the vacancy defect owing to the missing atom. For graphene, the atoms around the vacancy reconstruct in an asymmetric way, such that two of the three atoms around the vacancy get closer to each other relative to the third atom. Then two different second neighbor distances, 1.93 \AA~ and 2.55 \AA~ are distinguished. Hence, while the interatomic distances AB and AC in ~\ref{fig1} (a) remained very close to the second nearest neighbor distance of pristine graphene, the distance between B and C atoms is reduced and becomes comparable with the first nearest neighbor distance. At the end, one atom at the edge of the defect becomes two-fold coordinated while the rest of the atoms remain three-fold coordinated. This geometrical asymmetry also leads to an asymmetric distribution of magnetic moments in the lattice and the ground state of graphene with single vacancy acquires a net permanent magnetic moment. On the other hand, the atoms in silicene with single vacancy defect reorient themselves to display a three fold rotation symmetry; three atoms around the vacancy move towards the center of the defect and ground state becomes nonmagnetic. The interatomic distances AB, AC and BC are all equal to 2.74 \AA. They are much smaller than the second nearest neighbor distance(3.93 \AA~) of pristine silicene, but close to its first nearest neighbor  distance 2.27 \AA. Under these circumstances, A, B and C atoms can be viewed as fourfold coordinated despite two of their bonds are slightly longer than the remaining two. Nonetheless, both the heart type asymmetric reconstruction of graphene vacancy and the symmetric reconstruction of silicene vacancy create chemically active sites. In the next section, we will show that the interaction of adsorbates with graphene or silicene at the vacancy site acquires crucial features owing to these chemically active sites.

\subsection{Dissociative adsorption at the vacancy of graphene and silicene}

\begin{figure*}
\begin{center}
\includegraphics[width=14cm]{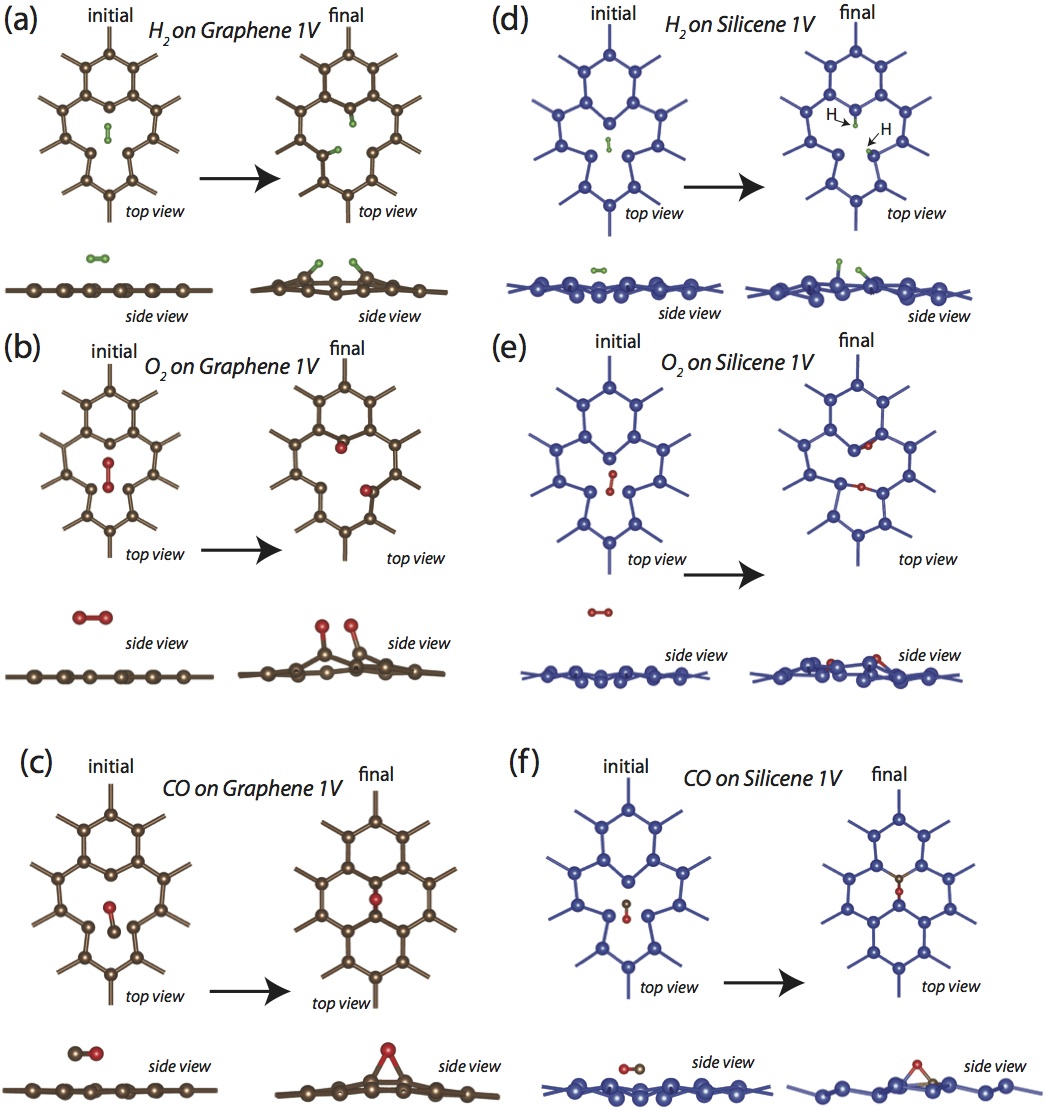}
\caption{Top and side views of the dissociative adsorption of (a) H$_2$ ,(b) O$_2$ and (c) CO at the vacancy site
of graphene. The constituent H, O and C atoms are bonded to the carbon atoms around the reconstructed vacancy. At the end the atomic configurations are modified. Top and side views of the dissociative adsorption of (d) H$_2$, (e) O$_2$ and (f) CO at the vacancy site of silicene. Upon dissociation, while constituent H atoms are bonded to Si atoms
around the vacancy, O atoms prefer to form two bridge bonds.  In the ball-stick model, H, O, and C atoms are
represented by green, red and brown spheres, respectively.}
\label{fig2}
\end{center}
\end{figure*}

Initial and final atomic configurations leading to the dissociative adsorption of H$_2$, O$_2$ and CO at the vacancy site of graphene and silicene are presented in ~\ref{fig2}. Hydrogen molecule is initially placed parallel to the graphene plane at various locations around the vacancy defect. It remains intact as long as the perpendicular distance between the molecule and the graphene layer is greater than $\sim$1.3 \AA~. Only a weak vdW interaction of 93 meV occurs between the molecule and graphene \cite{H2}. For weakly interacting H$_2$ and defected graphene, the magnetic moment is 1 $\mu_B$. However, if the molecule gets closer than $h_c$=1.3 \AA~ to graphene by overcoming a barrier of $E_{B}=$ 0.82 eV, an attractive interaction between the vacancy site and H atoms sets in and the molecule starts to be drawn towards the defect. As the molecule approaches to the plane of graphene, each H atom starts to move toward the dangling C atoms located around the defect. While the two-fold coordinated C atom is strongly bonded to graphene and remains undetached, one of hydrogen atoms of H$_2$ starts to get further away from the other H atom to form a new bond with this C atom. As for the second H atom, it is adsorbed to a second C atom surrounding the vacancy. In the final configuration, the distance between two hydrogen atoms increases to 1.47 \AA, which is more than twice the H-H bond length of 0.7 \AA~ in H$_2$ molecule. This way, the dissociation of H$_2$ molecule is achieved as shown in ~\ref{fig2}(a). The asymmetric arrangement around the reconstructed vacancy disappears, since the three-fold coordination of C atoms around the vacancy is approximately satisfied upon the saturation of dangling bonds by constituent H atoms of dissociated H$_2$. At the end, all the C-C distance around the vacancy become approximately equal. Under these circumstances, the net magnetic moment of the bare reconstructed vacancy diminishes. This is a remarkable effect of the dissociation of H$_2$, which should be detected experimentally.

While O$_2$ molecule is normally inert to pristine graphene, it is attracted by the vacancy site once it overcomes a small energy barrier of $E_{B}=$0.15 eV at about $h_c$=2.45 \AA~ above the plane of graphene. The O-O distance, that is normally 1.2 \AA~ in the O$_2$ molecule, increases to 1.6 \AA~ as it approaches to the vacancy. Eventually, O$_2$ dissociates and the constituent O atoms are bonded to C atoms around the defect as shown in ~\ref{fig2}(b). Again, the atomic configurations at the defect site is modified and consequently the ground state changes into the nonmagnetic state.

At about $h_{c} \sim$ 3 \AA~  above the surface, the attractive interaction between the defected graphene and CO positioned parallel to the surface is only $\sim$100 meV. At this position the magnetic moment of the defected graphene (1 $\mu_B$) is conserved. Once it overcomes an energy barrier at $h_c$=1.56 \AA, CO is attracted towards the vacancy by a relatively stronger chemical interaction and eventually is bonded from its carbon end to the two-fold coordinated C atom around the vacancy. If CO molecule is released from a height relatively closer to the surface (for example 1.50 \AA) CO dissociates. While the constituent C atom heals the vacancy by completing the perfect hexagon of C atoms, other constituent, namely O atom forms a bridge bond (C-O-C) above the two nearest neighbor C atoms of the hexagon as shown in ~\ref{fig2}(c). At the end of the dissociation process the magnetic moment of the defected graphene is diminished. Notably, due to its asymmetry the initial location and orientation of CO molecule relative to the surface is crucial for its dissociation or adsorption as a molecule at the vacancy site.

\begin{figure*}
\begin{center}
\includegraphics[width=16cm]{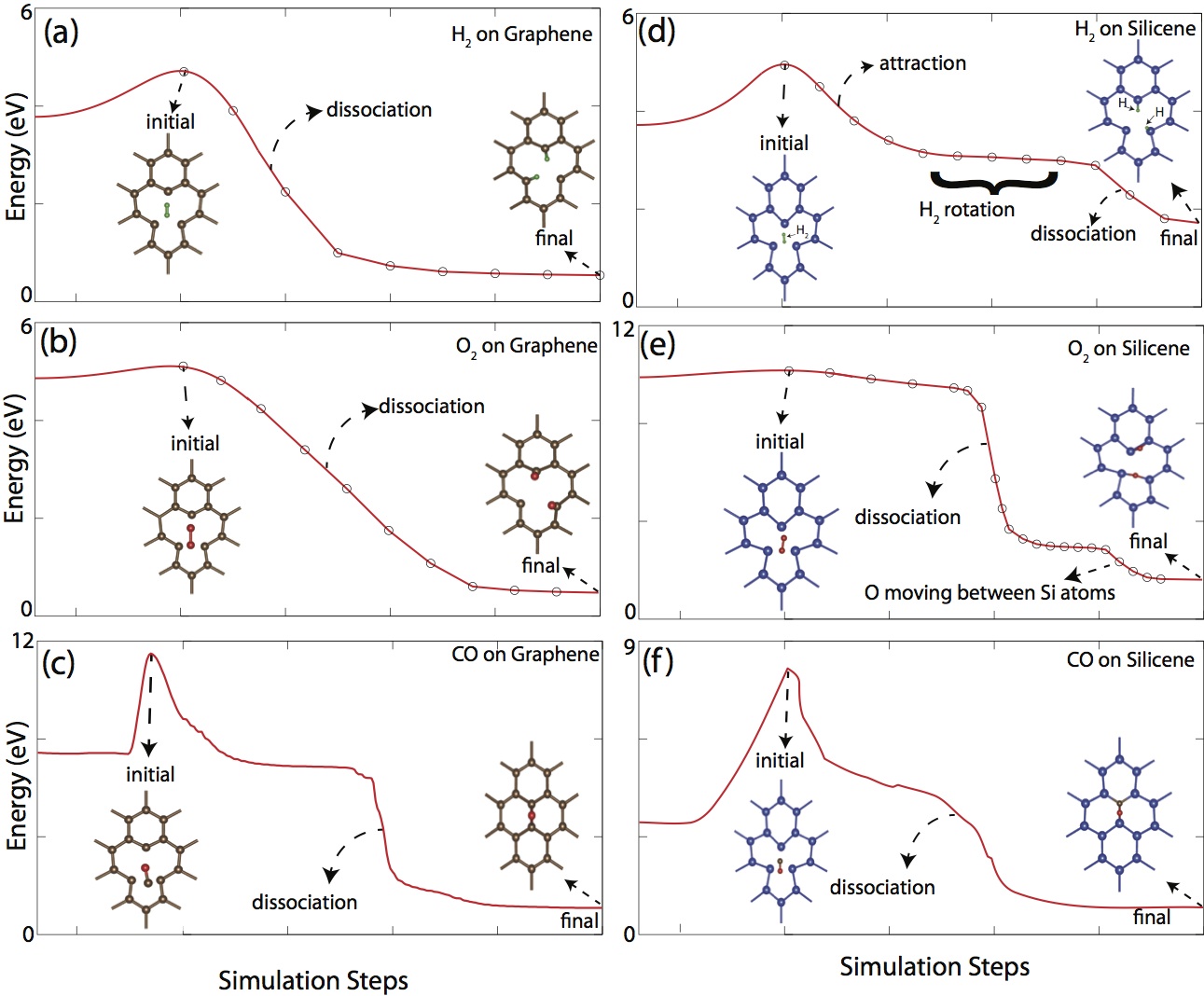}
\caption{Variation of the total energy at various steps during the dissociation process of (a) H$_2$ on graphene,
(b) O$_2$ on graphene,(c) CO on graphene (d) H$_2$ on silicene, (e) O$_2$ on silicene and (f) CO on silicene.}
\label{fig3}
\end{center}
\end{figure*}

Similar to graphene, H$_2$, O$_2$ and CO molecules dissociate readily at the vacancy sites of silicene, as shown in \ref{fig2}(d),(e) and (f).  The dissociation process of H$_2$ starts once it overcomes a barrier of $E_{B}=$1.3 eV at the critical height $h_{c}\simeq $ 1.1 \AA~ above the plane of silicene. Upon dissociation of H$_2$, two constituent H atoms are bonded to Si atoms at the edge of the vacancy from the top sites. In this respect, the bonding geometry is reminiscent of the dissociative adsorption of H$_2$ at the vacancy site of graphene. However, a different situation occurs when O$_2$ dissociates at the vacancy site of silicene. Firstly, the energy barrier is rather low ($E_B$=0.05eV)  for O$_2$ and the range of attractive interaction is rather long. Secondly, the constituent O atoms are adsorbed at two different bridge sites. This adsorption geometry is reminiscent of oxidized pristine silicene which leads to monolayer honeycomb structure of silicatene.\cite{silicatene} Finally, once overcame an energy barrier, CO molecule is also attracted towards the vacancy site of silicene. Carbon atom of the molecule tends to complete the hexagon having five Si atoms by substituting the missing Si atom. As for the constituent O atom of CO, it forms a bridge bond between the substituted C atom and nearest Si atom. This way the dissociation of CO is finalized as shown ~\ref{fig2}(f).

In ~\ref{fig3} the relevant stages in the course of the approaches of H$_2$, O$_2$ and CO from distant heights summarize energetics related with the dissociations of these molecules. For H$_2$ and O$_2$ at the vacancy site of graphene in ~\ref{fig3} (a) and (b), once a relatively small energy barriers are overcame, the total energies drop gradually as the molecules are dissociated. The dissociation processes are exothermic and the dissociation energies are $E_{d}=$4.82 eV and 4.95 eV, respectively. In ~\ref{fig3} (c), the energy barrier for the dissociation of CO is $E_{B}=$ 3.95 eV. Then the dissociation energy per CO molecule is $E_{d}=$6.26 eV. On the other hand, the energy associated with the molecular adsorption of CO to the two-fold coordinated carbon atoms at the vacancy site of graphene is $E_{a}=$1.78 eV.

Different stages are distinguished in the dissociation of H$_2$ and O$_2$ at the vacancy site of silicene in ~\ref{fig3} (d) and (e). For example, in the course of dissociation of H$_2$ at the vacancy site of silicene, the total energy first gradually decreases similar to that of the energy of H$_2$ on graphene.  Then comes a stage where energy remains almost constant in which the H$_2$ molecule aligns itself at the top of the missing bond of the silicene. During this period, H$_2$ doesn't move in the perpendicular direction but it rather rotates and moves in a horizontal plane. Once H$_2$ gets between the two two-fold coordinated Si atoms at the vacancy site, the total energy starts to drop once again and the dissociation of H$_2$ is completed. In this exothermic process the dissociation energy is calculated to be $E_{d}\simeq$ 2 eV.

The dissociation of O$_2$ at the vacancy site of silicene displays also different stages. In addition to the above stages seen in the dissociation of H$_2$, in the last stage two Si-O bonds reorient themselves to form two bridge bonds as seen in ~\ref{fig3}(e). This final step manifests itself in the energy plot as a stepped topology. The energy associated with the dissociation of O$_2$, which results in two Si-O-Si bridge bonds around the vacancy is $E_{d}$=8.54 eV. These bridge bonded oxygen atoms are common to the oxidation of pristine silicene, which appears to be the energetically most favorable absorption geometry.  It should be noted that O$_2$ is more prone to dissociate at the vacancy site as compared to H$_2$. This situation is attributed to the higher electronegativity of O (3.44) relative to that of H (2.20) according yo the Pauling scale. Oxygen atoms tend to form Si-O or C-O bonds rather than O-O bond. This property is reflected to the energy released after the dissociation process. We note that the oxidation of graphene is of current interest due to almost reversible oxidation-deoxidation process with continuous metal-insulator transition promising potential device applications \cite{topsakal}. In view of the fact that the interaction between O$_2$ molecule and graphene or silicene is weak and hence does not render dissociation on pristine structures, chemically active sites around the vacancy provide suitable medium to nucleate the process of oxidation. In the dissociation of CO at the defected site of silicene the barrier and dissociation energies are $E_{B}= 4.36$ and $E_{d}=2.42 eV$, respectively. In ~\ref{table1} we summarize the calculated values of critical energies and structure parameters associated with the dissociation or adsorption of H$_2$, O$_2$ and CO molecules. As a final remark of this section, we note that even if H$_2$ and O$_2$ are initially positioned parallel to graphene and silicene surfaces at different sites above the single vacancy, their orientations undergo a change as soon as they start to interact with the vacancy. However, the initial orientation of an asymmetric molecule like CO can be crucial for the final state.

\begin{table*}
\caption{ Calculated values of critical energies and structure parameters associated with the dissociation or adsorption of H$_2$, O$_2$ and CO molecules at the vacancy site of graphene and silicene: Dissociation / adsorption energy ($E_d$ / \textbf{$E_a$}); energy barrier for dissociation / adsorption ($E_B$ / \textbf{$E_A$}), critical dissociation distance ($h_c$), the distance of the physisorbed molecule ($h_p$) are presented. }
\label{table1}
\begin{center}
\begin{tabular}{ccccccccc}
\hline  \hline
& \multicolumn{2}{c}{$E_d$ / $E_a$)(eV)}  & \multicolumn{2}{c}{$E_B$ / ($E_A$)(eV)} & \multicolumn{2}{c}{$h_c$(\AA~)} & \multicolumn{2}{c}{$h_p$(\AA~)}\\
Molecule & Graphene & Silicene & Graphene & Silicene & Graphene & Silicene & Graphene & Silicene \\
\hline
$H_2$ & 4.82 & 2.00 & 0.82 & 1.30 & 1.30 & 1.10 & -      & - \\
$O_2$ & 4.95 & 8.54 & 0.15 & 0.05 & 2.45 & 5.00 & -      & -   \\
$CO$  & 6.26 / \textbf{(1.78)} & 2.42 & 3.95 / \textbf{(3.54)} & 4.36 & 1.50 & 1.70 & $\sim$ 3   & $\sim$ 4   \\
\hline
\hline
\end{tabular}
\end{center}
\end{table*}

\begin{figure*}
\begin{center}
\includegraphics[width=16cm]{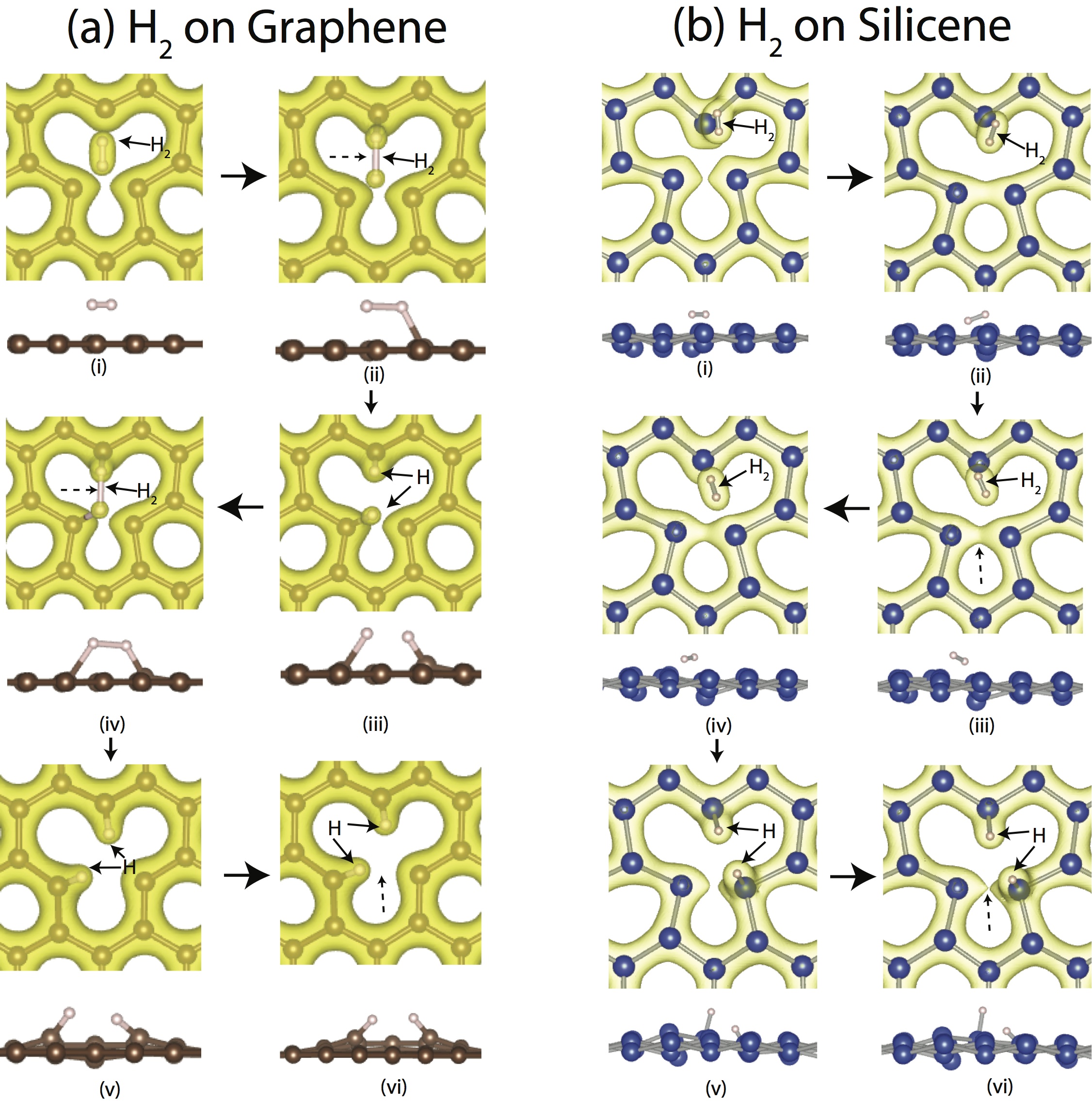}
\caption{Modifications of atomic configuration and variation of the isosurfaces of the total charge distribution in the course of the dissociation of H$_2$ molecule at the vacancy site of graphene (a) and silicene (b). The concerted actions of Si atoms at the close proximity of vacancy and that of H$_2$ molecule are revealed. Similar behaviors are also observed in other dissociation processes occurring at vacancy site of graphene. Specific bond weakening and bond forming are highlighted by dotted arrows.}
\label{fig4}
\end{center}
\end{figure*}

\subsection{Charge density analysis}
The total charge densities at the close proximity of the single vacancy defect together with the incoming molecule is examined in the course of dissociation. In ~\ref{fig4} we present the variation of the total charge distribution as H$_2$ molecule dissociates at the vacancy site of silicene. Note that similar trends are also observed in other dissociation processes examined in this study. Once an energy barrier is overcame and an attractive chemical interaction sets in between the H$_2$ molecule and silicene, the molecule starts to approach the defect site and concomitantly it reorients itself parallel to the missing bond. Subsequently, one of H atoms of H$_2$ is attracted first by one of the two fold coordinated Si atoms at the defect site to establish the Si-H bond. The charge transfer from H$_2$ to this Si-H bond weakens the H-H bond. Eventually, the second constituent H atom is drawn by another Si atom around the vacancy to form the second Si-H bond. This way, the dissociation of H$_2$ is completed. Meanwhile, the reconstructed configuration of the Si-Si bonds around the vacancy changes to be close to the ideal configuration. We note that the gradual changes in the charge density distribution and atomic configuration in the course of dissociation process show that the dissociation of molecules is attained by the \textit{concerted action} of atoms and bonds at the close proximity. This is why the energy barrier for dissociation is rather low.

\begin{figure*}
\begin{center}
\includegraphics[width=16cm]{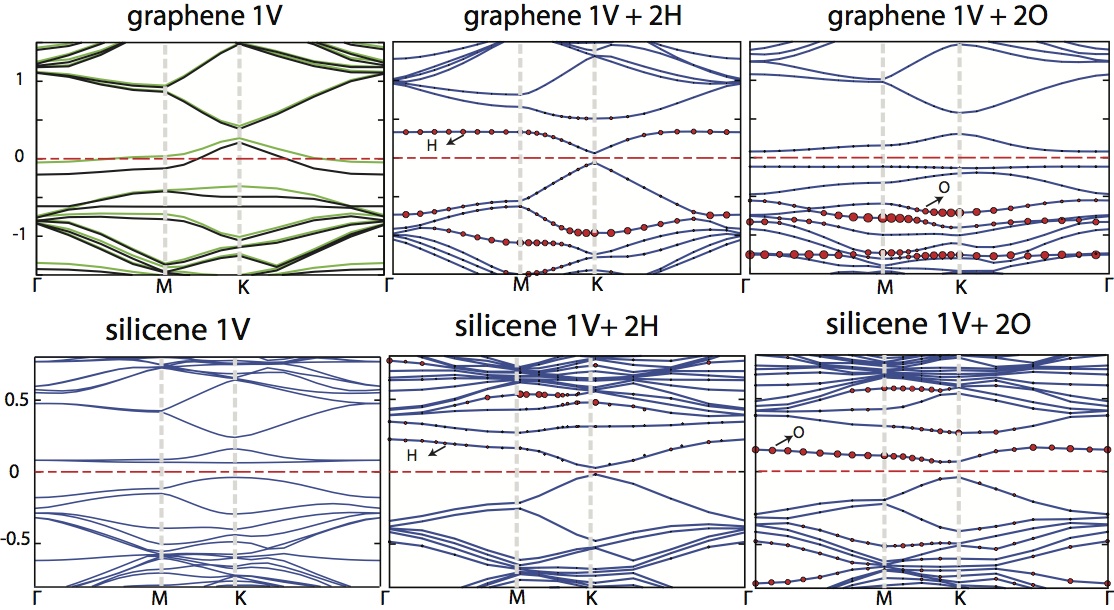}
\caption{Energy band structures  of graphene and silicene which have a single vacancy (1V) in each $(8 \times 8)$ supercell before and after the  dissociation of H$_2$ and O$_2$. Spin degenerate, spin up and spin down bands are shown by blue, black and green lines, respectively. The  zero of energy is set to the Fermi level. Upper panels of the figure show the energy band structure of graphene having a bare single vacancy (1V) and those modified upon the dissociative adsorption of H$_2$ and O$_2$. Owing to the magnetic ground state of the defected graphene, the bands are split. Lower panels of the figure show the energy band structure of silicene having a bare vacancy (1V) and those modified upon the dissociative adsorption of H$_2$ and O$_2$. The red dots indicate the contributions of H or O atoms to graphene/silicene band structures.}
\label{fig5}
\end{center}
\end{figure*}

\subsection{Electronic structure}
Electronic structures of graphene and silicene having a single vacancy (1V) in the $(8 \times 8)$ supercell before and after the dissociation of H$_2$ and O$_2$ are presented in \ref{fig5}. Because of supercell geometry, the translational symmetry is maintained and the states related with the vacancy generate energy bands $E(\vec{k})$ which are periodic in the reciprocal space. In reality, the vacancy gives rise to the localized (resonance) defect states in the band gaps (band continua) of graphene and silicene. Nevertheless, the coupling between defects treated within the supercell geometry is reduced and consequently the related energy bands are flattened as the distance between defects increases. In \ref{fig5}, the states related with vacancy are singled out for the low dispersion of their bands.

The states of the reconstructed bare vacancy of graphene show hole-like behavior. In the periodic structure of the  vacancy of graphene, linearly crossing bands split, but the structure becomes metallic and magnetic, whereas silicene with a reconstructed bare vacancy in each $(8 \times 8)$ supercell is nonmagnetic and semiconductor with a small band gap. The saturation of low coordinated host atoms around the vacancy of graphene and silicene by constituent H atoms following the dissociation of H$_2$ renders the structure semiconductor. Additionally, the magnetic moment of the reconstructed bare vacancy of graphene is diminished to zero. In the case of single, isolated vacancy defect, while bands of graphene and silicene continue to cross linearly at the Fermi level, the localized (resonance) states derived from C-H or Si-H bonds are located in the conduction and valence bands. Owing to the high electronegativity of the adsorbed O atoms, significant amount of charge is transferred from graphene to the adsorbed O atoms and the Fermi level is lowered. Because of the Si-O-Si bridge bonds, the resulting electronic structure formed following the dissociation of O$_2$ at the vacancy site of silicene displays a different band structure, where oxygen-related bands occur slightly above and below the Fermi level. The electronic structure following the dissociated CO on defected graphene, where the defect is healed and constituent O atom forms bridge bond is also presented.

\subsection{H$_2$O and OH molecules}
In contrast to H$_2$, O$_2$ and CO, the energy barriers for the dissociation H$_2$O and OH molecules are rather high and hence these molecules are prevented from dissociation into constituent atoms. Earlier it has been reported that water can dissociate at defective sites in graphene or carbon nanotubes following many possible reaction pathways some of which need to overcome high activation barriers.\cite{nardelli}. Instead of dissociation, OH molecule prefers to be adsorbed at the vacancy site of graphene through its oxygen end with a binding energy of $E_a$=4.21 eV. Notably, the binding energy of  OH to pristine graphene is only 0.97 eV.\cite{gurel} Similar to OH molecule, specific atoms or macro molecules, which are otherwise inert to pristine graphene and silicene can form rather strong chemical bonds at the vacancy site.

\section{Conclusions}
In conclusion, we revealed the increased chemical activity of host atoms with low coordination at the vacancy sites of graphene and silicene. Furthermore, we presented a detailed analysis of the effects of these chemically active sites on the adsorbed molecules. We showed that by overcoming a relatively small energy barrier, hydrogen and oxygen molecules can dissociate at the vacancy site. Upon dissociation, constituent H atoms saturate two-fold coordinated host atoms surrounding the vacancy and cause significant changes in electronic and magnetic properties. Although O$_2$ cannot be dissociated on pristine graphene and silicene, the vacancy site enables the dissociation and hence nucleates the oxidation of these single layer structures. The energy barrier for the dissociation of CO is relatively higher. However, the molecule is prone to dissociation once it overcomes this barrier. While constituent C atom tends to heal the vacancy defect, oxygen atom forms bridge bond. Our charge density analysis reveals the concerted action of host atoms together with adsorbed molecules in the course of dissociation, which lowers the potential barrier for dissociation.  Finally, we note that H$_2$O molecule is inert to graphene and silicene whereas OH molecule is adsorbed at the vacancy site with a significant binding energy. Present study indicates that graphene and silicene can be functionalized by creating meshes of single vacancy, where specific molecules can be either dissociated or pinned at the vacancy site. Resulting superstructures may render physical and chemical properties depending on the size and symmetry of meshes. In particular, biological macromolecules which are otherwise inert to pristine graphene and silicene can be bonded to the chemically active sites and eventually they become pinned to the surface. The superstructures produced by the meshes\cite{meshexp} of vacancies, where selected macro molecules are pinned can attain diverse functionalities for various technological applications.  We believe that the dissociations of molecules at the defected sites and the resulting changes in the electronic properties will promise further sensor applications.

\section*{Acknowledgements}
Part  of the computational resources has been provided by TUBITAK ULAKBIM, High Performance and Grid Computing Center (TR-Grid e-Infrastructure) and UYBHM at Istanbul Technical University through Grant No. 2-024-2007. The financial support of the Academy of Science of Turkey (TUBA) is acknowledged.


\begin{thebibliography}{99}

\bibitem{novoselov2004}
Novoselov, K. S.; Geim, A. K.; Morozov,S. V.; Jiang, D.;  Zhang, Y.;  Dubonos, S. V.; Grigorieva, I. V.; Firsov, A. A. Electric Field Effect in Atomically Thin Carbon Films. \emph{Science} \textbf{2004}, \emph{306}, 666-669.

\bibitem{geim2007}
Geim, A. K.; Novoselov, K. S. The Rise of Graphene. \emph{Nat. Mater.} \textbf{2007}, \emph{6}, 183-191.

\bibitem{zhang}
Li, H. ; Zhang, R. Vacancy-Defect-Induced Diminution of Thermal Conductivity in Silicene. \emph{EPL}, \textbf{2012}, \emph{99}, 36001-6.

\bibitem{seymur2009}
Cahangirov, S. ; Topsakal, M. ; Akturk, E.; Sahin, H. ; Ciraci, S.   Two- and One-Dimensional Honeycomb Structures of Silicon and Germanium.  \emph{Phys. Rev. Lett.} \textbf{2009}, \emph{102}, 236804-4.



\bibitem{engin}
Durgun, E.; Tongay, S.;  Ciraci, S.  Silicon and III-V Compound Nanotubes: Structural and Electronic Properties. \emph{Phys. Rev. B.} \textbf{2005}, \emph{72}, 075420-10.

\bibitem{nakamura}
Nakano, H.;  Mitsuoka, T.; Harada,  M.;  Horibuchi, K.;  Nozaki, H.;  Takahashi,N.;  Nonaka, T.;  Seno,Y.; Nakamura, H. Soft Synthesis of Single-Crystal Silicon Monolayer Sheets. \emph{Angew. Chem., Int. Ed.}, \textbf{2006}, \emph{45},  6303-06.

\bibitem{lay2010}
Aufray, B.; Kara, A.; Vizzini, S.; Oughaddou, H.; Leandri, C.; Ealet, B.; Le Lay, G. Graphene-Like Nanoribbons on Ag(110): A Possible Formation of Silicene. \emph{Appl. Phys. Lett.}, \textbf{2010}, \emph{96},  183102.

\bibitem{lay2012}
Vogt, P.;  De Padova, P.; Quaresima, C.; Avila, J.;  Frantzeskakis, E.; Asensio, M. C.; Resta, A.;  Ealet, B.;   Le Lay, G.  Silicene: Compelling Experimental Evidence for Graphenelike Two-Dimensional Silicon. \emph{Phys. Rev. Lett.} \textbf{2012}, \emph{108}, 155501-5.

\bibitem{seymur3} Cahangirov, S.; \" Oz\c celik, V. O.; Xian, L.; Avila, J.; Cho, S.; Asensio, M. C.; Ciraci, S.; Rubio A. Atomic Structure of the $ \sqrt{3} - \sqrt{3}$ Phase of Silicene on Ag(111) \emph{Phys. Rev. B.} \textbf{2014}, \emph{90}, 035448-4.

\bibitem{seymur4}
Cahangirov, S.; \" Oz\c celik, V. O.; Rubio, A.; Ciraci, S. Silicite: The layered allotrope of silicon, \emph{Phys. Rev. B.} \textbf{2014}, \emph{90}, 085426.

\bibitem{yazyev2010}
Yazyev, O. V.; Louie, S. G. Topological Defects in Graphene: Dislocations and Grain Boundaries. \emph{Phys. Rev. B.} \textbf{2010}, \emph{81}, 195420-7.

\bibitem{hashimoto2004}
Hashimoto, A. ; Suenaga, K.; Gloter, A.; Urita,K.; Iijima, S. Direct Evidence for Atomic Defects in Graphene Layers. \emph{Nature} \textbf{2004}, \emph{430}, 870-3.

\bibitem{krash2007}
Krasheninnikov, A. V.; Banhart, F. Engineering of Nanostructured Carbon Materials with Electron or Ion Beams. \emph{Nat. Mater.} \textbf{2007}, \emph{6}, 723.

\bibitem{crespi1996}
Crespi, V. H.;  Benedict, L. X.; Cohen, M. L.;  Louie, S. G.  Prediction of a Pure-Carbon Planar Covalent Metal. \emph{Phys. Rev. B} \textbf{1996}, \emph{53}, R13303.

\bibitem{kim2011}
 Kim, Y.;  Ihm, J.; Yoon, E.;   Lee, G. D.  Dynamics and Stability of Divacancy Defects in Graphene.\emph{ Phys. Rev. B} \textbf{2011}, \emph{84}, 075445-5.

\bibitem{singh2009}
Singh, R.;  Kroll, P. Magnetism in Graphene Due to Single-Atom Defects: Dependence on the Concentration and Packing Geometry of Defects. \emph{J. Phys.: Cond. Matter} \textbf{2009}, \emph{21}, 196002-7.

\bibitem{faccio2010}
 Faccio, R.; Werner, L. F.;  Pardo,H.;  Goyenola,C.;  Ventura, O. N.; Mombru, A. W.  Electronic and Structural Distortions in Graphene Induced by Carbon Vacancies and Boron Doping. \emph{J. Phys. Chem. C} \textbf{2010}, \emph{114}, 18961-18971.

\bibitem{neto2009}
Castro Neto, A. H.;  Peres, N. M. R.;  Novoselov, K. S.; Geim, A.   The Electronic Properties of Graphene. \emph{Rev. Mod. Phys.} \textbf{2009}, \emph{81}, 109-162.

\bibitem{reina2008}
Reina, A.;  Jia, X.; Ho, J.;  Nezich,D.;  Son, H.; Bulovic, V.;  Dresselhaus, M. S.;  Kong,  J. Large Area, Few-Layer Graphene Films on Arbitrary Substrates by Chemical Vapor Deposition. \emph{Nano. Lett.} \textbf{2008}, \emph{9},  30-35.

\bibitem{banhart2011}
Banhart, F.; Kotakoski, J.;  Krasheninnikov, A. V. Structural Defects in Graphene. \emph{ACS Nano} \textbf{2011}, \emph{5} , 26-41.

\bibitem{growth}
\" Oz\c celik, V. O.;  Cahangirov, S.; Ciraci, S. Epitaxial Growth Mechanisms of Graphene and Effects of Substrates. \emph{Phys. Rev. B} \textbf{2012}, \emph{85}, 235456.




\bibitem{zan2012}
 Zan, R.; Ramasse, Q. M.; Bangert, U. ;  Novoselov, K. S. Graphene Reknits Its Holes. \emph{Nano Lett.} \textbf{2012}, \emph{12}, 3936-3940.

\bibitem{robertson2013}
Robertson, A. W.; Montanari, B.; He, K.; Allen,C. S.;  Wu, Y. A.;  Harrison, N. M.;  Kirkland, A. I.;  Warner, J. H. Structural Reconstruction of the Graphene Monovacancy. \emph{ACS Nano} \textbf{2013}, \emph{7}, 4495-4502.

\bibitem{vozcelik2013}
\" Oz\c celik, V. O.;  Gurel, H. H.; Ciraci, S. Self-Healing of Vacancy Defects in Single-Layer Graphene and Silicene. \emph{Phys. Rev. B} \textbf{2013}, \emph{88}, 045440.

\bibitem{stone86}
Stone, A.; Wales, D. Theoretical Studies of Icosahedral C60 and Some Related Species. \emph{Chem. Phys. Lett.} \textbf{1986}, \emph{128}, 501.

\bibitem{wales98}
Wales, D. J.; Miller, M. A.;  Walsh, T. R. Archetypal Energy Landscapes,\emph{ Nature} \textbf{1998}, \emph{394}, 758.

\bibitem{wang}
Wang, C.; Ding, Y. Catalytically Healing the Stone-Wales Defects in Graphene by Carbon Adatoms \emph{J. Mater. Chem. A} \textbf{2013}, \emph{1} , 1885-1891.

\bibitem{boukhvalov}
Boukhvalov, D. V.; Katsnelson, M. I. Chemical Functionalization of Graphene with Defects, \emph{Nano Lett.} \textbf{2008}, \emph{8}, 4373-4379.

\bibitem{carlsson}
Carlsson, J. M.; Hanke, F; Linic, S.; Scheffler, M. Two-Step Mechanism for Low-Temperature Oxidation of Vacancies in Graphene, \emph{Phys. Rev. Lett.} \textbf{2009}, \emph{102}, 166104-4.


\bibitem{glukhova}
Glukhova, O.; Slepchenkov, M. Influence of the Curvature of Deformed Graphene Nanoribbons on Their Electronic and Adsorptive Properties: Theoretical Investigation Based on the Analysis of the Local Stress Field for an Atomic Grid, \emph{Nanoscale}, \textbf{2012}, \emph{4}, 3335-3344.

\bibitem{swnt}
Gulseren, O.; Yildirim, T.; Ciraci, S. Tunable Adsorption on Carbon Nanotubes, \emph{Phys. Rev. Lett.} \textbf{2001}, \emph{87}, 116802.

\bibitem{lin2007}
 Xu, S. C.; Irle, S. ; Musaev, D. G.; Lin, M. C. Sublimation of Ammonium Salts: A Novel Mechanism Revealed by a First-Principles Study of the $NH_4Cl$ System. \emph{J. Phys. Chem. C}, \textbf{2007}, \emph{111}, 1355-65.












\bibitem{lin2011}
Xu, S. C.; Chen, H. L.; Lin, M. C.  Band Gap Tuning of Graphene by Adsorption of Aromatic Molecules. \emph{J. Phys. Chem. C}, \textbf{2012}, \emph{116}, 1841-49.

\bibitem{can}
Ataca, C.;  Ciraci, S. Dissociation of $H_2O$ at the Vacancies of Single-Layer $MoS_2$. \emph{Phys. Rev. B}, \textbf{2012}, \emph{85}, 195410-6.

\bibitem{nardelli}
Kostov, M. K.;  Santiso, E. E.; George, A. M.;  Gubbins, K. E.;  Buongiorno-Nardelli, M. Dissociation of Water on Defective Carbon Substrates. \emph{Phys. Rev. Lett.}  \textbf{2005}, \emph{95}, 136105-9.


\bibitem{jiang2013}
Jiang, Q. G.;  Ao, Z. M.; Zheng, W. T.;  Li, S.;  Jiang, Q. Enhanced Hydrogen Sensing Properties of Graphene by Introducing a Mono-Atomic Vacancy. \emph{Phys. Chem. Chem. Phys.}  \textbf{2013}, \emph{15}, 21016.

\bibitem{sefa}
Dag, S.; Ozturk, Y.; Ciraci, S.;  Yildirim, T.  Adsorption and Dissociation of Hydrogen Molecules on Bare and Functionalized Carbon Nanotubes. \emph{Phys. Rev B.} \textbf{2005}, \emph{72}, 155404-8.

\bibitem{taner}
Yildirim, T.; Ciraci, S. Titanium-Decorated Carbon Nanotubes as a Potential High-Capacity Hydrogen Storage Medium .\emph{Phys. Rev. Lett.} \textbf{2005}, \emph{94}, 175501-4.

\bibitem{engin2}
Durgun, E. ; Ciraci, S. ; Zhou, W.; Yildirim, T.   Transition-Metal-Ethylene Complexes as High-Capacity Hydrogen-Storage Media. \emph{Phys. Rev. Lett.} \textbf{2006}, \emph{97}, 226102-4.

\bibitem{can2}
Ataca, C.; Akturk, E.; Ciraci, S.; Ustunel, H.  High-Capacity Hydrogen Storage by Metalized Graphene. \emph{Appl. Phys. Lett}. \textbf{2008}, \emph{93}, 043123.

\bibitem{yao}
Liu, C. C.; Feng,  W.; Yao, Y. Quantum Spin Hall Effect in Silicene and Two-Dimensional Germanium. \emph{Phys. Rev. Lett.}, \textbf{2011},  \emph{107},  076802.

\bibitem{dikin}
Dikin, D. A.; Stankovich, S.; Zimney, E. J.; Piner, R.D.; Dommett, G. H.; Evmenenko, G.; Nguyen, S. T.; Ruoff, R. S. Preparation and Characterization of Graphene Oxide Paper. \emph{Nature} \textbf{2007}, \emph{448}, 457-60.

\bibitem{wei}
Wei, Z. et al. Nanoscale Tunable Reduction of Graphene Oxide for Graphene Electronics. \emph{Science} \textbf{2010}, \emph{328}, 1373-1376.

\bibitem{topsakal}
Topsakal, M; Gurel, H. H.; Ciraci, S. Effects of Charging and Electric Field on Graphene Oxide, \emph{J. Phys. Chem. C.} \textbf{2013}, \emph{117}, 5943.

\bibitem{silicatene}
\" Oz\c celik, V. O.;  Cahangirov, S.; Ciraci, S.  Stable Single-Layer Honeycomb Structure of Silica. \emph{Phys. Rev. Lett.} \textbf{2014}, \emph{112}, 246803.

\bibitem{tanveer}
Hussain, T; Panigrahi, P; Ahuja, R. Sensing Propensity of a Defected Graphane Sheet Towards CO, H$_2$O and O$_2$, \emph{Nanotechnology} \textbf{2014}, \emph{25}, 325501-6.

\bibitem{sef1}
Tongay, S.; Zhou, J.; Ataca, C.; Liu, J.; Kang, J. S.; Matthews, T. S.; You, L.; Li, J.; Grossman, J. C.; Wu, J. Broad-range Modulation of Light Emission in Two-dimensional Semiconductor by Molecular Physisorption Gating, \emph{Nano Lett.} \textbf{2013}, \emph{13}, 2831-2836.

\bibitem{sef2}
Tongay, S.; Zhou, J.; Ataca, C.; Fan, W.; Luce, A.; Kang, J. S.; Liu, J.; Ko, C.; Raghunathanan, R.; Zhou, J.; Ogletree, F.; Li, J.; Grossman, J. C.; Wu, J. Defects Activated Photoluminescence in Two-dimensional Semiconductors: Interplay Between Bound, Charged, and Free Excitons, \emph{Sci. Reports} \textbf{2013}, \emph{3}, 2657.





\bibitem{grimme06}
Grimme, S.  Semiempirical GGA-Type Density Functional Constructed with a Long-Range Dispersion Correction. \emph{J. Comput. Chem.} \textbf{2006}, \emph{27}, 1787-99.

\bibitem{paw}
Blochl, P. E.  Projector Augmented-Wave Method. \emph{Phys. Rev. B} \textbf{1994}, \emph{50}, 17953-27.

\bibitem{pbe}
Perdew, J. P.; Burke, K.; Ernzerhof, M. Generalized Gradient Approximation Made Simple. \emph{Phys. Rev. Lett.} \textbf{1996}, \emph{77}, 3865-3868.

\bibitem{MP}
Monkhorst, H. J.; Pack, J. D. Special Points for Brillouin-Zone Integrations. \emph{Phys. Rev. B} \textbf{1976}, \emph{13}, 5188.

\bibitem{vasp}
Kresse, G.; Furthmuller, J.   Efficient Iterative Schemes for Ab-initio Total-Energy Calculations Using a Plane-Wave Basis Set. \emph{Phys. Rev. B} \textbf{1996}, \emph{54}, 11169-11186.

\bibitem{H2}
Vosko, S. H.; Wilk, L.; Nusair,  M.  Accurate Spin-Dependent Electron Liquid Correlation Energies for Local Spin Density Calculations: A Critical Analysis. \emph{Can. J. Phys.} \textbf{1980}, \emph{58}, 1200.




\bibitem{gurel} Gurel,  H. H.;  Ciraci, S.  Enhanced Reduction of Graphene Oxide by Means of Charging and Electric Fields Applied to Hydroxyl Groups. \emph{J. Phys. Condens. Mat.}  \textbf{2013}, \emph{25}, 300507.


\bibitem{meshexp}
Bai, J.; Zhong, X.; Jiang, S.;  Huang, Y.; Duan, X. Graphene Nanomesh. \emph{Nat. Nanotech.} \textbf{2010}, \emph{5}, 190-194.





\end{thebibliography}
\end{document}